# Room-Temperature Skyrmion Thermopower in $Fe_3Sn_2$

*Qianheng Du, Myung-Geun Han, Yu Liu, Weijun Ren[†], Yimei Zhu, and Cedomir Petrovic\**

Qianheng Du, Myung-Geun Han, Yu Liu, Weijun Ren, Yimei Zhu and C. Petrovic
Condensed Matter Physics and Materials Science Department, Brookhaven National Laboratory, Upton 11973 New York USA
E-mail: petrovic@bnl.gov and qdu@bnl.gov
Qianheng Du and C. Petrovic
Materials Science and Chemical Engineering Department, Stony Brook University, Stony Brook 11790 New York USA
[†] Present address: Shenyang National Laboratory for Materials Science, Institute of Metal Research, Chinese Academy of Sciences, Shenyang, China

**Abstract**

We present first room-temperature thermoelectric signature of the skyrmion lattice. This was observed in $Fe_3Sn_2$, a Kagome Dirac crystal with massive Dirac fermions that features high-temperature skyrmion phase. The room-temperature skyrmion lattice shows magnetic-field dependence of the wavevector whereas thermopower is dominated by the electronic diffusion mechanism, allowing for the skyrmionic bubble lattice detection. Our results pave the way for the future skyrmion-based devices based on the manipulation of the thermal gradient.

**Introduction**

Strong electronic correlations and topology are widely recognized as fundamental sources of novel states of matter [1]-[5] and technologically important material properties. [6],[7] Nanoscale magnetic skyrmions in spin textures of chiral magnets are quintessential embodiment of this concept. [8],[9]

Magnetic skyrmions are commonly observed by microscopy techniques in the real space and neutron scattering in the reciprocal space. [10],[11] Experiments available far below room temperature reveal no discernible [12] or relatively small changes [13] in magnetic-field dependent thermoelectric properties in the course of transition to the skyrmion crystal.

In magnetic metals thermopower includes electronic diffusion, phonon or magnon-drag thermopower. [14] Thermoelectric signature of the skyrmion lattice, in particular, is of interest in spintronics and spincaloritronics for information processing. [15],[16] Ferromagnetic $Fe_3Sn_2$ with a geometrically frustrated kagome bilayer of Fe attracts considerable interest due to its magnetic structure, anomalous Hall effect (AHE), Dirac electronic states and room-temperature skyrmion lattice. [17]-[21] Previous studies also show that skyrmion lattice in $Fe_3Sn_2$ can be manipulated by spatially geometric confinement. [22] Moreover, single-chain skyrmion bubbles in 600 nm nanostripes were reported to be stable far above the room temperature, up to 630 K, thus making significant progress towards nanoscale skyrmion -based spintronic.[23] On the other hand, it is also of interest to manipulate skyrmionic textures by thermal gradients in magnetic nanodevices.[24]-[26] In this paper we show first evidence of the room-temperature skyrmion detection by thermopower in $Fe_3Sn_2$ and discuss relevant mechanism.

**Results and discussions**

**Results**

The powder XRD pattern of $Fe_3Sn_2$ shows that all observed peaks can be well fitted with the *R-3mh* space group [**Fig. 1(a)**] confirming high purity of the single crystals. The determined lattice parameters a = b =5.345(2) Å and c =19.780(2) Å are in good agreement with the reported values.[27] In the single-crystal XRD [**Fig. 1(b)**], only (00l) peaks are detected, indicating that the crystal surface is parallel to the hexagoal plane and orthogonal to the c axis. **Fig. 1(c)** shows the unit cell of $Fe_3Sn_2$ under the High-angle annular dark-field (HAADF) scanning transmission electron microscopy (STEM), showing no evidence for atomic defects.

In **Fig. 2** holographically reconstructed magnetization map shows hexagonally packed skyrmionic bubble lattice at the room temperature.[19] Interestingly, the helicity (in-plane spin rotation sense) is either clockwise or anticlockwise, yielding +1 and -1, respectively, as topological charges.

We performed real-space imaging of magnetic spin structures in the ab plane and their evolution under external magnetic field along the c-axis in transmission electron microscope at room temperature and we also also show the magnetic-field-dependent thermopower (S) in the well-established skyrmionic bubble phase [19] (**Fig. 3**). The Lorentz contrasts of skyrmionic bubbles [**Fig. 3(a-d)**] are similar to the previous Lorentz microscopy study.[19] We mapped out the projected in-plane magnetization by off-axis electron holography under the residual magnetic field (11.7 mT) at room temperature. In order to stabilize the skyrmionic bubbles at 11.7 mT, a large external magnetic field ~ 1000 mT was abruptly turned off. Under the residual magnetic field (~ 11.7 mT with the objective lense fully off) in our microscope, Lorentz microscopy image [**Fig. 3(a)**] shows coexistence of stripes and bubbles in the ab plane. The hexagonally packed skyrmionic bubble lattices are induced in external magnetic field in the range of 600 mT ~ 800 mT [**Fig. 3(b-d)**], corresponding to the magnetic field range where the anomaly is observed in the measurements of room-temperature thermoelectric power and heat capacity [**Fig. 3(e,f)**]. In addition, we note that the skyrmionic bubble lattice were completely annihilated with magnetic field above 1 T [19], which is consistent not only with the thermopower [**Fig. 3(e)**] but also heat capacity measurement [**Fig. 3(f)**].

From the data presented above it is clear that magnetic skyrmions, topologically protected nanoscale spin textures [28], show clear signature in room-temperature macroscopic thermal measurements. In what follows we discuss relevant microscopic mechanim. We first note that in $\mu_0 H$ = 9 T applied along the c-axis, changes in resistivity $\rho(T)$, thermal conductivity $\kappa$ and thermopower S are subtle [**Fig. 4(a,b)**]. Metallic $\rho(T)$ argues in favor of considerable electronic diffusion S(T) but the phonon-drag mechanism should also be considered. Thermopower changes from negative to positive at 124(1) K in the absence of magnetic field on cooling, however the sign change moves to 129(1) K and 130(1) K when 9 T is applied in the hexagonal (ab) plane and along the c-axis, respectively. As shown by the red dash dot line in the **Fig. 4(b)**, thermopower shows linear dependence on temperature above 124 K consistent with electronic diffusion. In metallic materials with substantial carrier density Lorentz force will affect thermal and electrical transport alike, whereas dominant carriers are often either electrons or holes.[29],[30] When the magnetic field is applied along the c-axis, resistivity is either unchanged or somewhat decreased above about 120 K and shows positive magnetoresistance (MR) below 120 K whereas there is a up to 20% decrease in κ(9 T) when compared to the κ(0 T) below 100 K, suggesting that electronic contribution to thermal conductivity is not negligible.

Low-temperature heat capacity offers further insight [**Figure 4(c)**]. From the fits [**Fig. 4(c) inset**] using the Debye model $C_v = \frac{12\pi^4}{5}R(\frac{T}{\theta_D})^3 + \gamma T$, the Debye temperature ($\theta_D$) and Sommerfeld coefficient ($\gamma$) are obtained. The values are $\theta_D = 237.0 \pm 0.6$ K and $\gamma = 2.003(5) \times 10^{-2}$ J/molK$^2$. The phonon velocity is ~2010 ms$^{-1}$.[31] Both electron and phonon part of heat capacity are calculated up to the room temperature and are also shown in **Figure 4(c)**.

Next, we evaluate phonon drag vs. electronic diffusion mechanism on thermopower. The characteristic peak in thermal conductivity observed on cooling [**Fig. 4(a)**] is phonon-related and it commonly arises due to competition between the point-defect/boundary scattering and the Umklapp phonon scattering mechanism.[32] Possible phonon drag effects are supported by the sign change of S(T) at 124 K [**Fig. 4(b)**]. If we take change in the band structure and effective mass of Fe$_3$Sn$_2$ into consideration, the low-temperature sign change can not be explained by electron diffusion within the framework of Mott formula: $S = -\frac{\pi^2 k_B^2 T}{3e}[\frac{1}{\sigma}\frac{d\sigma(\varepsilon)}{d\varepsilon}]_{\varepsilon=\varepsilon_F}$.[33]-[35] As a qualitative estimation, if Drude's formula $\sigma = \frac{ne^2\tau}{m^*}$ is adopted for the conductivity σ where n is carrier concentration, m* is effective mass and τ is relaxation time inversely proportional to the density of states, then the energy dependencies from the charge carrier density and τ are approximately balanced out, i.e. σ has the same energy dependence as $1/m^*$. For $Fe_3Sn_2$, around the Fermi energy the effective mass decreases.[21] This means σ will increase with energy and yield the negative sign of S. Consequently contributions from other scattering processes such as phonon-drag or magnon-drag must be taken into consideration. Fermi surface of $Fe_3Sn_2$ features two dominant electron pockets.[20],[21] Negative thermopower should be expected if the electronic diffusion part $S_d$ in $S = S_d + S_p$ prevails over phonon-drag contribution $S_p$. This is indeed observed above 124 (K) and is in agreement with a decrease in the absolute values of the temperature-dependent thermopower in 9 T when compared to the S(0 T) [**Fig. 4(b)**]. As we show below, electronic diffusion mechanism can explain the linear change of S with temperature above 124fK, but not the positive thermopower at lower temperatures.

In order to study the sign change, we plot only measured S(T) below 124 K [**Fig. 4(d)**], red solid circles]. For a single parabolic band system, the $S_d$ can be defined by the equation $\frac{\pi^2 k_B^2 T}{3eE_F}$.[36] By linear fitting the data above 124 K [**Fig. 4(b)**] and extrapolating it down to 2 K, we obtain the diffusive part of thermopower at low temperature. The fitted value of $E_F$ is 0.25(1)eV. This is an order of magnitude less than what is expected in metals, but is also in agreement with carrier concentration in $Fe_3Sn_2$ n ~ $10^{22}cm^{-3}$.[20] Phonon scattering part of thermopower $S_p$ [**Fig 4(d)** blue open circles] obtained by subtracting $S_d$ from the measured S clearly indicates positive contribution, resulting in thermopower sign change and net positive S values. It should be noted that magnon-drag $S_m = AT + BT^{\frac{3}{2}}$ [37] could also influence thermopower in $Fe_3Sn_2$. In **Fig. 4(d)**, green line shows the fitting of low temperature $S_p$ using the power law $S = AT + BT^n$. The fitted n is 2.67±0.06 with A = 0.182 ± 0.002 and B = -(1.9 ± 0.6) ×$10^{-4}$. Whereas magnon-drag contribution cannot be completely excluded, it is evident that the fitted exponent is closer to $T^3$ dependence, expected in the phonon-drag mechanism of thermopower. The phonon mean free path (MFP) which is closely related to the phonon transport can be calculated by Fourier's law ($\kappa_P = \frac{1}{3}Cvl_\kappa$) with the Debye model, where $\kappa_p$ is phonon thermal conductivity obtained from by subtracting electronic thermal conductivity $\kappa_e$ from measured $\kappa(T)$. The $\kappa_e$ is estimated from the Wiedemann-Franz law $\kappa_e/T = L_0/T$, where $L_0 = 2.45 \times 10^{-8}$ W Ω K$^{-2}$ and ρ is the measured resistivity. The C, v and $l_\kappa$ are the phonon specific heat [**Fig. 4(c)**], phonon velocity and MFP of the phonon, respectively. The results are shown in the inset of **Fig. 4(d)**. Whereas phonon-drag mechanism is commonly associated with much longer mean-free path, we note that phonon drag in metals may not vary significantly with mean free path if energies of electron and phonon distributions are well matched, i.e. if the probability of electron interaction with quasi-balistic phonons is proportional to the size of the region where phonons propagate without mutual collisions.[38],[39] The presence of phonon contribution to thermopower explains the magnetic-field induced changes at low temperature. The absolute value of $S_d$ decreases in 9 T due to the Lorentz force whereas the $S_p$ remains unchanged. Since $S_d$ and $S_p$ have the opposite sign and at low temperature $S_p$ dominates, the net thermopower will increase. This explains the increase of thermopower crossover temperature from 124 K to 129 K for magnetic field in the hexagonal plane or 130 K when the magnetic field is applied along the easy magnetization c-axis.

**Discussion**

The above discussion confirms that, whereas phonon or magnon drag contribute to low-temperature thermopower, electronic diffusion mechanism is dominant at the room-temperature. In **Fig. 3 (e)** the magnetic field is applied along the c-axis and perpendicular to the thermal current flow, so that the thermal transport is measured in the plane of the skyrmionic lattice. When a thermal gradient $\nabla T$ is applied, a diffusion electric current density $J_d$ is generated. The diffusion part of thermopower $S_d$ is the ratio between the electric field required to stop $J_d$ and the $\nabla T$.[40] When a conduction electron transverses a skyrmion, it is affected by the local magnetization and hence continuously changes direction acquiring Berry phase.[41],[42]

Consequently, carriers experience effective Lorentz force which increases the Hall effect. Similar effects should take place when carriers are driven by external thermal gradient.[15] Indeed, as shown in **Fig.3(e)**, thermopower shows well-defined anomaly in the skyrmionic bubble phase. This is in contrast to thermopower outside the skyrmion region, such as for example in the spin glass state at 50 K. When electronic system enters skyrmion region, there is an increase in the absolute values of S(B). For systems with spin degree of freedom, the entropy is composed by the entropy from spin and crystal lattice. Ordering spin texture in skyrmionic phase will decrease the spin entropy and transfer it to the crystal lattice.[43] This part of entropy gives extra driving force to the thermal diffusion of conduction electrons, which will enhance the thermopower as magnetic stripe domains are gradually transformed into bubbles with increasing Zeeman field.[19] Further increase in magnetic field up to 9 T generates additional negative effect on thermopower from the spontaneous magnetization that exceeds the contributions from the entropy, resulting in a decrease in S as skyrmion bubble spin textures die out in higher magnetic fields. It also should be noted that magnetic-field dependent heat capacity [**Fig. 3(f)**] shows small, but discernible changes at room temperature. The oscillation amplitudes are up to about 2%, comparable to thermopower-related oscillations at 28 K in MnSi, and are absent outside (H,T) skyrmion region [**Fig. 3(e)**]. Since both thermopower S and heat capacity C show magnetic field-induced anomalies where skyrmions form,[19] this also may suggest that thermopower variations stem from topological quantum oscillations induced by DOS changes [44] where Zeeman field increase in skyrmion crystal induces DOS oscillations that arise from the magnetic-field dependence of the wavevector Q and directly affect electronic thermopower from the Mott formula relation.[44] Interestingly, two independent sets of triple-q systems under 585.1 mT were observed in our Lorentz microscopy, as shown in **Fig. 3(c)**. With further increasing magnetic field to 727.5 mT [**Fig. 3(d)**], one set (q' ) disappears and the other set (q) becomes dominant in the sample, showing a clear magnetic-field dependence of the wave vector in the skyrmionic bubble lattice.

**Summary and conclusion**

In summary, we present first signature of room-temperature skyrmion spin textures by thermopower. Therrmal transport in the high-temperature region is governed by electronic diffusion mechanism, enabling detection of topologically protected skyrmionic spin textures. Our results open new possibilities for skyrmion manipulation in future information storage and spin caloritronic devices using thermal gradients.[45],[46]

**Experimental Section**

*Crystal synthesis*

Single crystals of $Fe_3Sn_2$ were grown using flux method.[47] Whereas some crystals were initially grown by mixing Fe and Sn in 5:95 stoichiometry, heating to 1150 °C, holding at this temperature for 24 hours, fast-cooling to 910 °C and then slow cooling to 800 °C [48], cooling to 770 °C was used to to increase the size of crystal to about 3 mm length.[20]

*Characterization*

Crystal structure was determined by analyzing powder x-ray diffraction (XRD) pattern taken with Cu K$\alpha$($\lambda$ = 0.15418 nm) radiation of Rigaku Miniex powder di_ractometer. Transmission electron microscopy (TEM) samples are prepared by focused ion beam (FIB) using 5 keV Ga+ ions for a final milling (FEI Helios 600). The range of the collection angle used for High-angle annular dark-field (HAADF) scanning transmission electron microscopy (STEM) was 68 - 280 mrad. The HAADF STEM image was filtered using a Fourier mask in the Digital Migraph software (Gatan, Inc.). Aberration-corrected JEOL ARM 200CF and JEOL 2100F Lorentz were used for Lorentz imaging and off-axis electron holography, respectively, at the 200 keV operation voltage. The external magnetic field was applied by controlling excitation of the objective lens.

*Electrical and thermal transport measurements*

Resistivity, heat capacity and thermal transport (TTO) properties were measured in a Quantum Design PPMS-9. Standard four contact method was used to measure transport properties. Electrical resistivity ($\rho$) was measured for the current ow in the hexagonal plane and the magnetic field (H) was applied along the c axis whereas thermal conductivity ($\kappa$) and thermopower S measurements were taken for the heat flow along the ab plane and magnetic field applied both in the hexagonal plane and along the c axis of the crystallographic unit cell.

**Acknowledgements**

This work was funded by the Computation Material Science Program (Y. L.and C. P.). TEM studies at Brookhaven National Laboratory were supported by US DOE, Office of Science, Office of Basic Energy Sciences under contract DE-SC0012704. Weijun Ren thanks the Study Abroad Program of the China Scholarship Council for support. Hologram reconstruction and analysis were performed with custom scripts written by Martha R. McCartney (Arizona State University).

Received: ((will be filled in by the editorial staff))
Revised: ((will be filled in by the editorial staff))
Published online: ((will be filled in by the editorial staff))

References

[1] D.S.L. Abergel, V. Apalkov, J. Berashevich, K. Ziegler and T. Chakraborty, Adv. Phys. **2010**, 59, 261.

[2] X. Wan, A. M. Turner, A. Vishwanath and S. Y. Savrasov Phys. Rev. B **2011**, 83, 205101.

[3] B. Yang, E. Moon, H. Isobe and N. Nagaosa, Nat. Phys. **2014**, 10, 774.

[4] S. A. Yang, H. Pan, and F. Zhang, Phys. Rev. Lett. **2014**, 113, 046401.

[5] H. Wei, S. Chao, and V. Aji Phys. Rev. B **2014**, 89, 235109.

[6] S. D. Huber, Nat. Phys. **2016**, 12, 621.

[7] C. Beenakker and L. Kouwenhoven, Nat. Phys. **2016**, 12, 618.

[8] T. Skyrme, Nucl. Phys. **1962**, 31, 556.

[9] N. Bogdanov and D. A. Yablonskii, Sov. Phys. JETP **1989**, 68, 101.

[10] X. Z. Yu, Y. Onose, N. Kanazawa, J. H. Park, J. H. Han, Y. Matsui, N. Nagaosa and Y. Tokura, Nature **2010**, 465, 901.

[11] S. Mühlbauer, B. Binz, F. Jonietz, A. Rosch, A. Neubauer, R. Georgii and P. Böni, Science **2009**, 323, 915.

[12] Y. Hirokane, Y. Tomioka, Y. Imai, A. Maeda and Y. Onose, Phys. Rev. B **2016**

93, 014436.

[13] S. Arsenijević, C. Petrovic, L. Forró, and A. Akrap, Euro. Phys. Lett. **2013**, 103, 57015.

[14] S. J. Watzman, R. A. Duine, Y. Tserkovnyak, S. R. Boona, H. Jin, A. Prakash, Y. Zheng, and J. P. Heremans, Phys. Rev. B **2016**, 94, 144407.

[15] A. Hoffmann and Sam D. Bader, Phys. Rev. Applied **2015**, 4, 047001.

[16] G. E. W. Bauer, Eiji Saitoh and Bart J. van Wees, Nat. Mater. **2012**, 11, 391.

[17] L. A. Fenner, A. A. Dee, and A. S. Wills, J. Phys.: Condens. Matter **2009**, 21, 452202.

[18] T. Kida, L. A. Fenner, A. A. Dee, I. Terasaki, M. Hagiwara, and A. S. Wills, J. Phys.: Condens Matter **2011**, 23, 112205.

[19] Z. Hou, W. Ren, B. Ding, G. Xu, Y. Wang, B. Yang, Q. Zhang, Y. Zhang, E. Liu, F. Xu, W. Wang, G. Wu, X. Zhang, B. Shen, and Z. Zhang, Adv. Mater. **2017**, 29, 1701144.

[20] Q. Wang, S. Sun, X. Zhang, F. Pang, and H. Lei, Phys. Rev. B **2016**, 94, 075135.

[21] L. Ye, M. Kang, J. Liu, F. V. Cube, C. R. Wicker, T. Suzuki, C. Jozwiak, A. Bostwick, E. Rotenberg, D. C. Bell, L. Fu, R. Comin, and J. G. Checkelsky, Nature **2018**, 555, 638.

[22] Z. Hou, Q. Zhang, G. Xu, S. Zhang, C. Gong, B. Ding, H. Li, F. Xu, Y. Yao, E. Liu, G. Wu, X. Zhang, and W. Wang, ACS Nano, 2019, 13, 922.

[23] Z. Hou, Q. Zhang, G. Xu, C. Gong, B. Ding, Y. Wang, E. Liu, F. Xu, H. Zhang, Y. Yao, G. Wu, X. Zhang, and W. Wang, Nano Lett. **2018**, 138, 1274.

[24] U. Martens, T. Huebner, H. Ulrichs, O. Reimer, T. Kusche, R. R. Tamming, C.-L. Chang, R. I. Tobey, A. Thomas, M. Münzenberg, J. Walowski, Commun. Phys. **2018**, 1, 65.

[25] J. C. Le Breton, S. Sharma, H. Saito, S. Yuasa, R. Jansen, Nature **2011**, 475, 82.

[26] D. J. Kim, C. Y. Jeon, J. G. Choi, J. W. Lee, S. Surabhi, J. R. Jeong, K. J. Lee, B. G. Park, Nat. Commun. **2017**, 8, 1400.

[27] B. Malaman, B.Roques, A. Curtois and J. Protas, Acta Crystallogr. B **1976**, 32, 1348.

[28] Y. Z. Liu, and J. Zang, Acta Physica Sinica **2018**, 67, 131292.

[29] C. Uher, Thermal conductivity of metals, in Thermal Conductivity: Theory, Properties, and Applications; Tritt, T. M., Eds.; Kluwer Academic Press, 2004.

[30] P. G. Klemens and R. K. Williams, Int. Mater. Rev. **1986**, 31, 197.

[31] O. Andersen, J. Phys. Chem. Solids **1963**, 24, 909.

[32] J. Yang, Theory of Thermal Conductivity, in Thermal Conductivity: Theory, Properties and Applications; Tritt, T. M., Eds.; Kluwer Academic Press, 2004.

[33] M. Cutler and N. F. Mott, Phys. Rev. **1969**, 181, 1336.

[34] B. Xu and M. J. Verstraete, Phys. Rev. Lett. **2014**, 112, 196603.

[35] L. Ye, M. K. Chan, R. D. McDonald, D. Graf, M. Kang, J. Liu, T. Suzuki, R. Comin, L. Fu and J. G. Checkelsky, Nat. Commun. **2019**, 10, 4870.

[36] M. Kargarian and G. A. Fiete, Phys. Rev. B **2013**, 88, 205141.

[37] F. J. Blatt, D. J. Flood, V. Rowe, and P. A. Schroeder, Phys. Rev. Lett. **1967**, 18, 395.

[38] E. D. Eidelmann, and A. Y. Vul, J. Phys.: Condens. Matter **2007**, 19, 266210.

[39] A. Sergeev, A. Mitin, A. Phys. Rev. B **2001**, 65, 064301.

[40] D. G. Cantrell and P. N. Butcher, J. Phys. C: Solid State Phys. **1987**, 20, 1985.

[41] J. Ye, Y. B. Kim, A. J. Millis, B. I. Shraiman, P. Majumdar, and Z. Tesanovic, Phys. Rev. Lett. **1999**, 83, 3737.

[42] C. Franz, F. Freimuth, A. Bauer, R. Ritz, C. Schnarr, C. Duviange, T. Adams, S. Blügel, A. Rosch, Y. Mokrousov and C. Peiderer, Phys. Rev. Lett. **2014**, 112, 186601.

[43] J. R. Gómez, R. F. Garcia, A. D. M. Catoira, and M. R. Gómez, Renew. Sustain. Energy Rev. **2013**, 17, 74.

[44] S. Sorn, S. Divic, A. Paramekanti, Phys. Rev. B **2019**, 100, 174411.

[45] N. Nagaosa, Y. Tokura, Nat. Nanotechnol. **2013**, 8, 899.

[46] X. G. Wang, L. Chotorlishvilli, G. H. Guo, C. L. Jia, and J. Berakdar, Phys. Rev. B **2019**, 99, 064426.

[47] Z. Fisk, J. P. Remeika, Chpater 81 Growth of single crystals from molten metal fluxes,

in Handbook on the Physics and Chemistry of Rare Earths; Gschneidner, K. A. Jr; Eyring, L., Eds.; Vol. 12, pp 53-70 Elsevier, Amsterdam, 1989.

[48] W. Ren, and C. Petrovic (unpublished).

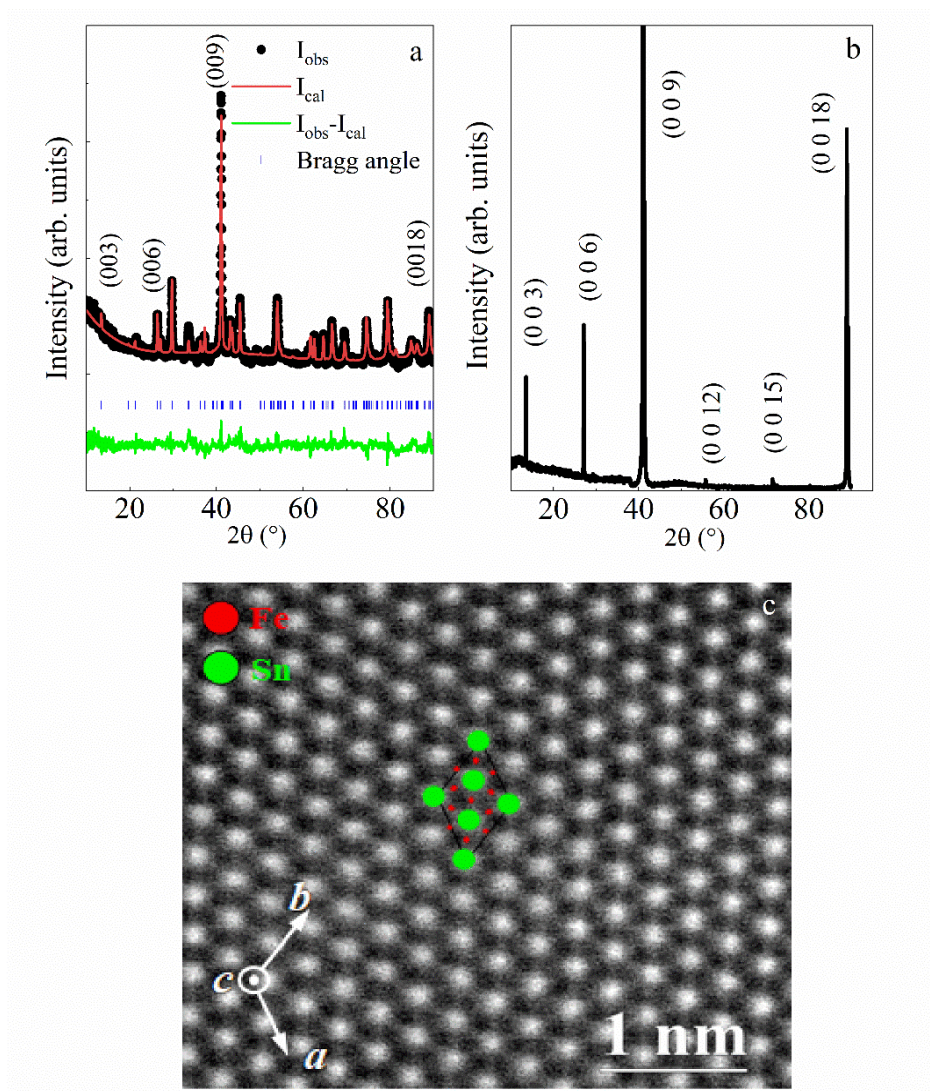

**Fig. 1** (a) Powder x-ray diffraction (XRD) and (b) single crystal 2θ scans of $Fe_3Sn_2$ at room temperature. The vertical tick marks in (a) represent Bragg reections of the R-3mh space group. High-angle annular dark-field (HAADF) scanning transmission electron microscopy (STEM) image of $Fe_3Sn_2$ taken along the c-axis. Fe columns are somewhat less visible as the heavier Sn columns have two times more atoms that that of Fe columns in addition to large atomic number difference. A unit cell model is embedded in the image.

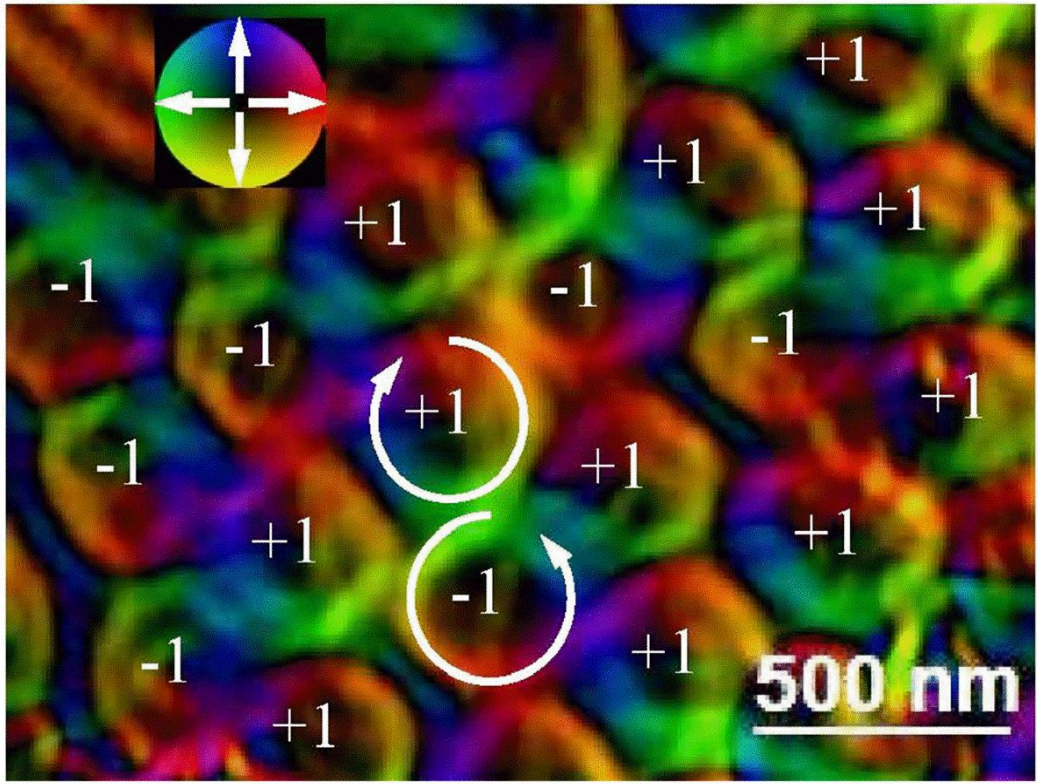

**Fig. 2** Color-contour composite image obtained from the magnetic phase image reconstructed from the off-axis electron hologram. The electron hologram was taken with external magnetic field 11.7 mT. The skyrmionic bubbles are induced by rapidly changing magnetic field from 1000 mT to 11.7 mT. Based on the in-plane spin rotation sense, the topological charge of skyrmionic bubble is determined as ±1.

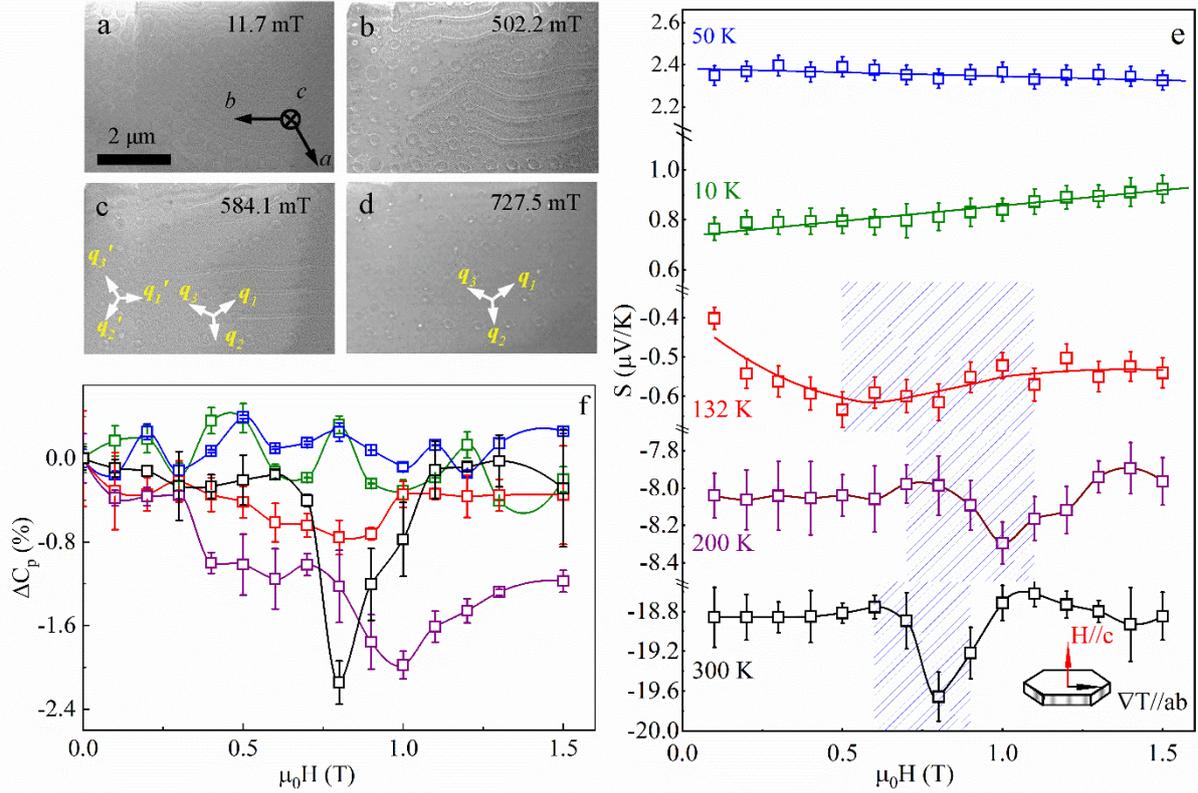

**Fig. 3** (a-d) Lorentz images showing the magnetic field dependence of the hexagonal plane spin textures in $Fe_3Sn_2$. The defocus value was about + 500 µm. The arrows show the crystallographic directions determined from the electron diffraction (not shown here). The external magnetic field was applied along the imaging direction using the objective lens coil. There are two independent sets of triple-q systems in 584.1 mT (c). With further increasing magnetic field to 727.5 mT (d), one set (q' ) disappears and the other set (q) becomes dominant in the sample. (e) Thermopower vs magnetic field at several temperatures near where the skyrmionic phase exists. The same data at 10 and 50 K are also plotted for comparison. The shadow area indicates the skyrmionic phase. Note that the temperature of the best-defined thermopower anomalies in skyrmion phase corresponds well to temperature where maximum density of skyrmion bubles was observed [Ref. 19].(f) Relative change of heat capacity in magnetic field at the identical temperature; legend in (e) also denotes temperatures in (f).

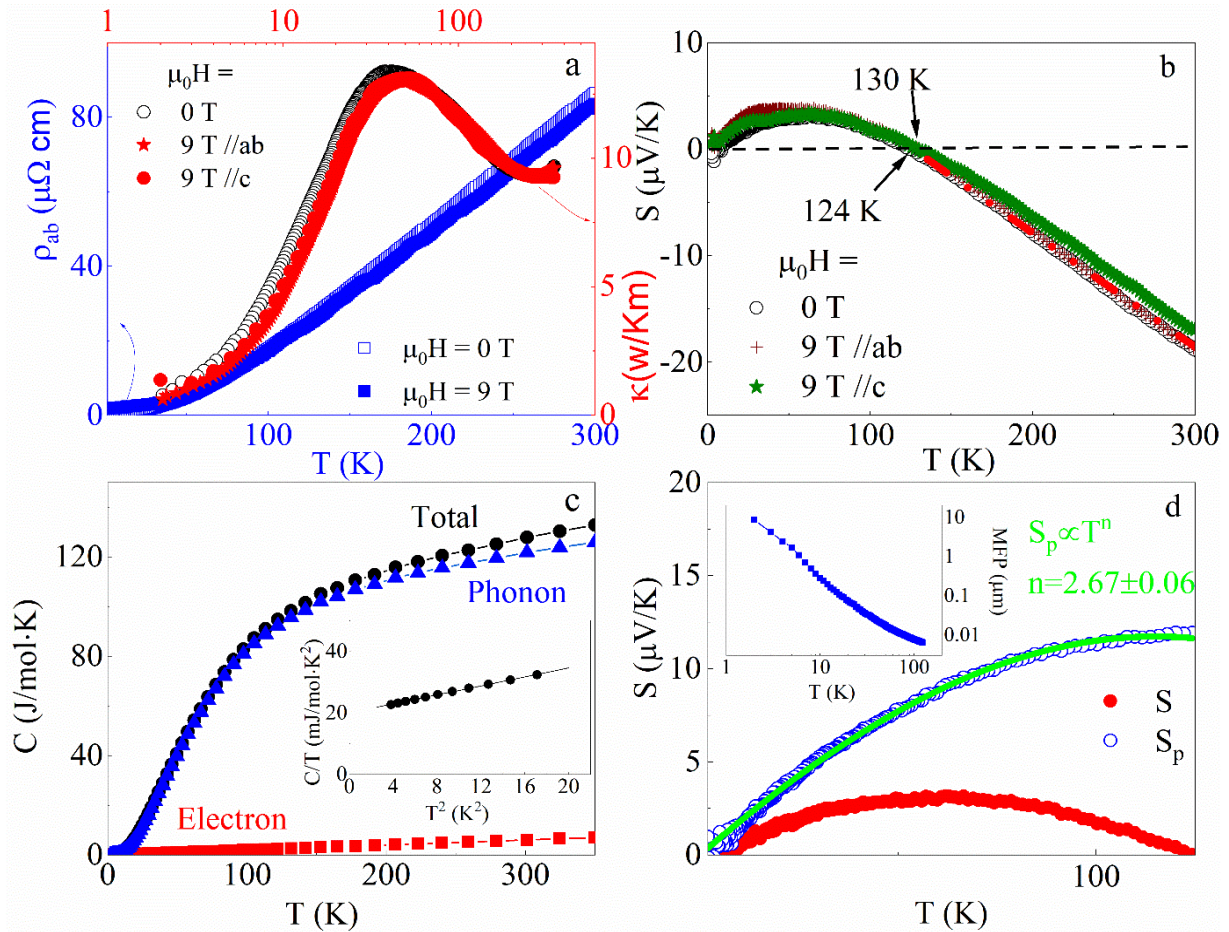

**Fig. 4** (a) Temperature dependence of electrical resistivity and thermal conductivity. (b) Temperature dependence of thermopower. The blue cubic shows zero field data and the red circle shows data in 9 T field along ab plane. The green star shows data in 9 T field along c axis. Note linear S(T) dependence at high temperatures indicated by red dash dot line. (c) Temperature dependence of heat capacity. (d) Temperature dependence of positive part of thermopower. The red circles are the measured thermopower below the 124 K. The blue open circles represent the $S_p$ extracted. The green line shows the power law fitting of $S_p$. The inset shows the phonon MFP below 124 K.

**The table of contents entry. Field-dependence** of thermopower at 300 K with field and temperature gradient along c axis and ab plan respectively. When the skyrmion bubbles form, there is a drop in the value of thermopower. The inset shows the skyrmion bubbles under 727.5 mT field.

**Keyword**: thermoelectrics, magnetic materials, skyrmion

Qianheng Du, Myung-Geun Han, Yu Liu, Weijun Ren, Yimei Zhu, and Cedomir Petrovic

**Room-Temperature Skyrmion Thermopower in Fe$_3$Sn$_2$**

ToC figure

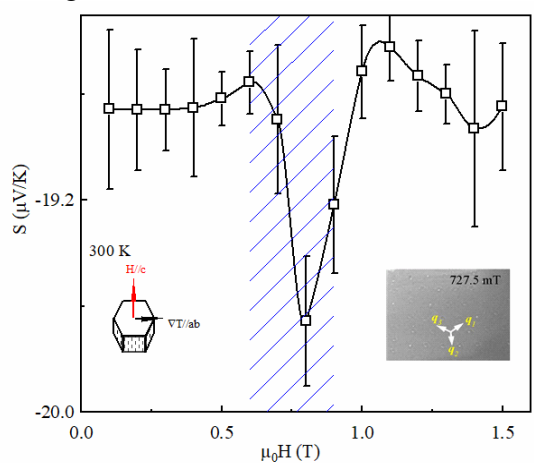